# Turbostratic graphitic microstructures: electronically decoupled multilayer graphene devices with robust high charge carrier mobility

*Yenny R. Hernandez[1,2], Sebastian Schweitzer[3], June-Seo Kim[2], Ajit Kumar Patra[3], Jan Englert[4], Ingo Lieberwirth[1], Andrea Liscio[5], Vincenzo Palermo[5], Xinliang Feng[1], Andreas Hirsch[4], Mathias Kläui[*2,3] and Klaus Müllen[*1]*

[1]Max Planck Institute for Polymer Research, Ackermannweg 10, 55128 Mainz, Germany.

[2]Institut für Physik, Johannes Gutenberg-Universität Mainz, Staudingerweg 7, 55128 Mainz, Germany. [3]Universität Konstanz, Fachbereich Physik, 78457 Konstanz, Germany. [4]Friedrich Alexander University Erlangen –Nürmberg, Henkestrasse 42, 91054 Erlangen, Germany. [5]ISOF – Istituto Sintesi Organica a Fotoreattivitá CNR, Area della Ricerca di Bologna, Via Gobetti 101, 40129 Bologna, Italy.

## Abstract

Carbon nanomaterials continue to amaze scientists due to their exceptional physical properties[1]. Recently there have been theoretical predictions and first reports on graphene multilayers, where, due to the rotation of the stacked layers, outstanding electronic properties are retained while the susceptibility to degradation and mechanical stress is strongly reduced due to the multilayer nature[2]. Here we show that fully turbostratic multilayer graphitic microstructures combine the high charge carrier mobilities necessary for advanced electronic and spintronic devices[3] with the robustness of graphitic structures. Structural characterization of disk-shaped graphitic microstructures using Raman spectroscopy and Transmission Electron Microscopy (TEM) reveals Moiré and diffraction patterns corroborating their turbostratic nature. Electronic transport characterization yields reproducible high mobilities >$10^5$ cm$^2$V$^{-1}$s$^{-1}$ independent of the disks' thickness, which is a direct consequence of the electronic decoupling induced by the turbostratic stacking.

The high temperature growth of carbon nanostructures has led to the discovery of allotropes such as fullerenes[4], carbon nanotubes (CNT)[5], carbon cones and carbon disks[6]. The latter are composed of turbostratically stacked graphene layers[7]. In particular, graphene multilayers grown on SiC have been shown to exhibit a nearly ideal band structure[8] and monolayer-like Raman features[9]. These features have been ascribed to the electronic decoupling induced by the rotation of adjacent layers[10], theoretically predicted to occur for rotation angles between 1.47° and 30° [2a]. Here we study multilayer disks that can be grown in large quantities by the pyrolysis of hydrocarbons in a plasma torch process. We show that their robust fully turbostratic nature results in reproducibly high charge carrier mobilities as required for future electronic and spintronic devices and comparable to suspended graphene.

The graphitic disks were dispersed in 1-methyl-2-pyrrolidinone using bath sonication followed by centrifugation allowing for the separation of the disks from each other and from other types of microstructures (for more details see supplementary information Figure S1 (a)). High resolution

TEM (HRTEM) of the disks' edges (disks on holey carbon grids, Figure 1 (a)) revealed curving of the outer layers (Figure 1 (b)). This curving effect is not continuous around the disk (Figure 1 (b) inset), which means that the layers are independent of each other. Parallel beam nano-diffraction patterns of the disks are composed of many diffraction spots that are attributed to the rotation of the stacked layers (Figure 1 (c)). We find Moiré interference patterns (Figure 1 (d)) as previously observed on (i) epitaxially grown graphene multilayers[2b], (ii) rotated layers in mechanically cleaved few-layer graphene[11] and (iii) few-layer graphene structures produced by chemical means[12]. The observed patterns are in line with those theoretically derived[2a] for turbostratic graphene with layer rotations of θ ≥ 5º. We find that the rotation angle varies between disks (Figure S3), resembling the situation of CNT grown under similar conditions where it is not possible to control the nanotube chirality[1b]. The rotation angle between layers is directly related to the periodicity of the observed Moiré pattern:

$$D = \frac{a}{2\sin(\theta/2)} \qquad (1)$$

with D the superlattice periodicity, $a$ the basal lattice constant (0.246 nm for HOPG) and θ is the rotation angle between two layers of the hexagonal lattice. The periodicity of the Moiré pattern in Figure 1 (d) (~2 nm) yields a rotation angle of ~7º, which is corroborated by measuring the angular difference between diffraction spots (Figure 1 (c)).

In Figure 2 (a) we plot representative Raman spectra of a turbostratic disk and natural graphite. We observe similar spectra for all disks analyzed suggesting that the Raman spectrum does not depend on the disk thickness or the stacking angle. The shape and intensity of the 2D peak is significantly different from what has been reported for graphite or multilayer graphene. In the case of CNT the 2D peak appears as a single peak with an intensity comparable to the intensity of the G peak. In our case Raman mapping yields $I_{2D}/I_G \sim 0.75$ (peak intensity -Figure S4) which is in line with the CNT case. We can fit the 2D peak with three Lorentzian functions separated by ~3 meV. Such a fit of the 2D peak with multiple Lorentzians has been attributed to the presence of van-Hove singularities (vHs) in CNT[13], which previously have also been observed in twisted bilayer graphene[14] and predicted to induce magnetic states[14].

As turbostratic stacking of graphene multilayers is predicted to lead to electronic decoupling, resulting in monolayer-like electric transport properties[2c], we studied this by contacting disks in a van der Pauw-configuration (vdP)[15] (Figure 3 (a) inset) (for more details see supplementary information S.I.2.). Figure 3 (a) shows the sheet resistance ($R_s$) of several disks plotted as a function of their thicknesses (t) at low temperature (LT = 4.3 K). Clear inverse scaling is observed and fitting with $R_s = \rho/t$, yields a resistivity $\rho = 3.52 \pm 0.11 \times 10^{-6}$ $\Omega$m. From the observed scaling we conclude that virtually all of the layers participate in the charge transport. The resistivity of the devices is constant for all numbers of stacked layers: $\rho = (3.45 \pm 0.42) \times 10^{-6}$ $\Omega$m (Figure 3 (b)), in good agreement with the fit. Moreover, this demonstrates a high reproducibility of the disks' electrical properties in contrast to single layer graphene, which can exhibit very different $R_s$ even on nominally identical substrates. The temperature dependence of $R_s$ shows a reproducible decrease from LT to room temperature (RT) of more than a factor of 2 (Figure 3 (b) inset). This observed resistance drop with temperature (see also supplementary information) suggests that the interior of the disks is comprised of layers with low carrier concentrations[16].

Given these indications for low charge carrier concentrations, the low $R_s$ must stem primarily from large charge carrier mobilities. To corroborate this, we plot in Figure 3 (c) the Hall values obtained for different disks at different temperatures. At LT the observed signals cover a wide range of negative and positive values. At RT all disks show a reproducible positive Hall signal. This can be explained by assuming two types of layers: First, surface layers, doped extrinsically by impurities, adsorbates or charge puddles in the substrate[17] and second, inner screened layers, which do not sense any extrinsic doping. For the doped surface layers which typically show high charge carrier concentrations already at LT ($\sim 10^{12}$ cm$^{-2}$), carrier concentrations do not change significantly over the temperature range. For the inner screened layers however, the charge carrier concentration significantly increases at higher temperature and dominates the transport signal at RT: Therefore, the observed uniform positive Hall voltages at RT are caused by a change in the charge carrier concentrations at the inner layers and point to an overall hole dominated transport.

If this explanation holds, the back-gate dependence of the resistance should be strongly temperature dependent as it acts only on the bottom interface layers. We measure the back-gate dependence of the resistance to calculate the charge carrier densities ($n_s$) and mobilities ($\mu$), for both surface and inner layers. The disks show an ambipolar electric field effect with a maximum in resistivity and an almost symmetric drop with applied back-gate voltage ($V_{BG}$) (Figure 3 (d)). This behavior is known from graphene[16a] resulting from the symmetric band structure around the Dirac points (conductivity $\sigma$ proportional to $n_s$). The maximum of the resistance at the neutrality point, and the difference R(0 $V_{BG}$) - R($\pm$100 $V_{BG}$) feature a reciprocal drop with temperature (see Figure S6).

To further analyze the observed ambipolar effect, we use a screening length of $\lambda = 1.2 \pm 0.2$ nm as reported from measurements on top- and bottom-gated thin films of graphite[18] which is expected to be similar for turbostratic graphene. With an interlayer spacing of 0.34 nm, the induced charge carrier concentration ($i$) in layer i, counted from the bottom layer i=0, is

$$n_i \approx \frac{C_{SiO2}}{e}(V_{BG} - V_{Dirac})e^{-i/3.5} \quad (2)$$

where $C_{SiO2}$ is the capacitance of the SiO$_2$ (300 nm) and $V_{Dirac}$ is the voltage to reach the Dirac point. This implies a contribution to the signal of the back-gate measurements of about ~ 4 layers for our stack comprised of ~ 60 layers. The additionally introduced charge carrier density is then given by $n \approx 4 \times n_0 \approx 3 \times 10^{13} \ cm^{-2}$. The maximum of the resistance can be achieved by an applied V$_{BG}$ of ~ ±10 V. This means, by introducing ±3 × 10$^{12}$ cm$^{-2}$ charge carriers, the observed resistances can be maximized. At this point, the bottom layers, which are influenced by the substrate, are tuned to the Dirac point. Thus the observed Hall voltages, at $V_{BG} = 0$, are caused by extrinsic doping. The overall concentration of the inner layers has to be in any case smaller than half of the induced charge carrier concentration at 10 V, meaning: $n_{Bulk} \approx 56 \times nb \lesssim$ 1.5 × 10$^{12}$ $cm^{-2}$, yielding for each inner layer $nb \lesssim 10^{10} \ cm^{-2}$. At the peak resistances, where $R_S$ ~ 26 Ω (at LT), the bottom layers are close to the neutrality point (NP). Then the total resistance is caused by ~ 56 (= 60 - 4) inner layers, with each of them having a carrier concentration of ~ $10^{10} \ cm^{-2}$. Therefore, with 26 Ω ≈ 1/ (56×10$^{10}$×μ×1.6×10$^{-19}$) Ω, each inner layer has to exhibit a mobility of $\gtrsim 10^5 \ cm^2V^{-1}s^{-1}$. This value shows that the inner layer mobility in our structures is competitive with the highest mobilities found in suspended graphene[17] while retaining the robust properties of a multilayer structure.

In summary, we study fully turbostratic graphitic multilayer microstructures that combine the easy processability, robustness and high charge carrier mobilities necessary for advanced devices. TEM characterization of isolated graphitic disks reveals Moiré and diffraction patterns that confirm their fully turbostratic nature. The transport characterization of many disks yields reproducible mobilities of >$10^5$ cm$^2$V$^{-1}$s$^{-1}$ independent of the thickness of the disks. While the screening of the inner layers prevents effective gating, the high charge carrier mobility qualifies turbostratic graphene for a range of devices including interconnects and given the scaling of the spin diffusion length with charge mobility[19], for spintronics devices[3]. In particular, as recently shown for few layer epitaxial turbostratic graphene, huge spin diffusion lengths can be found in these materials[3]. This can be explained by the large charge carrier mobility that we found, making our material suited for efficient spin transport.

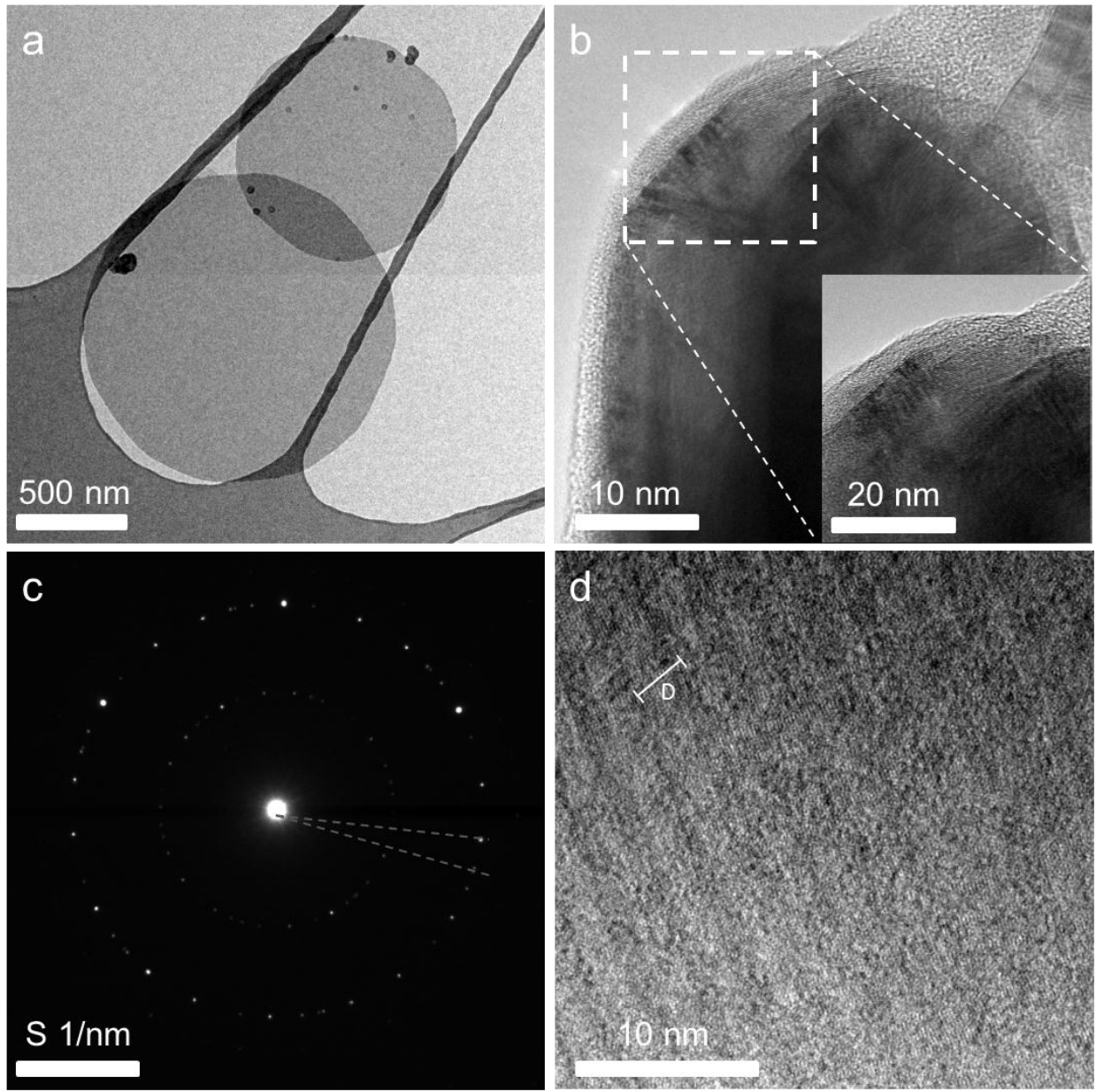

**Figure 1.** Turbostratic disks (a) TEM image of disks dispersed on a holey carbon grid. (b) Cross section image of the edge of a carbon disk showing curving of the edges (c) Parallel beam nano-diffraction taken at the surface of the disk composed by several diffraction spots that are due to contributions of non-Bernal stacked layers. (d) High resolution microscopy of the disk shows Moiré patterns. The Moiré periodicity, D, can be measured to be ~2 nm in this case.

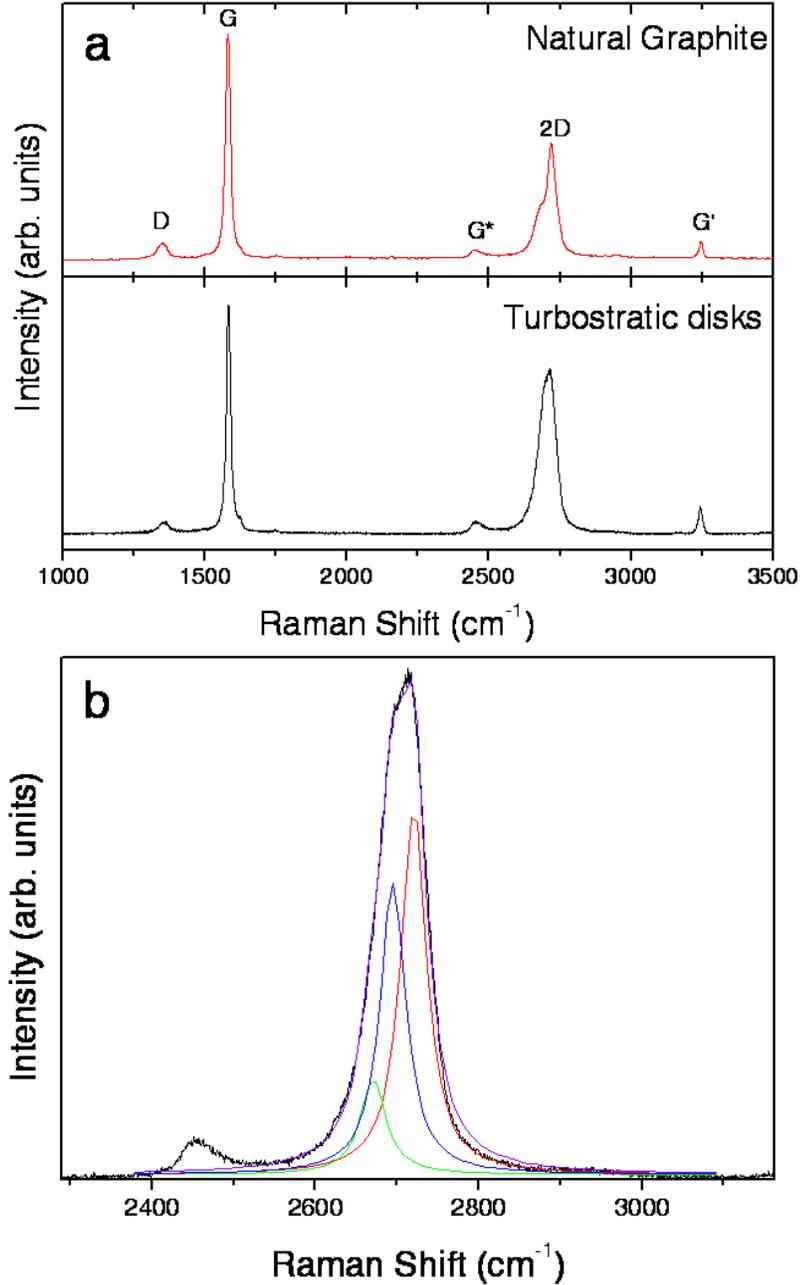

**Figure 2.** Raman spectroscopy of single carbon disks deposited on 300 nm SiO2 (a) Typical Raman spectra for natural graphite and for a single carbon disk recorded at 532 nm excitation wavelengths. (b) The 2D peak observed for all the disks analyzed can be fitted by 3 Lorentzian curves separated by ~3 meV.

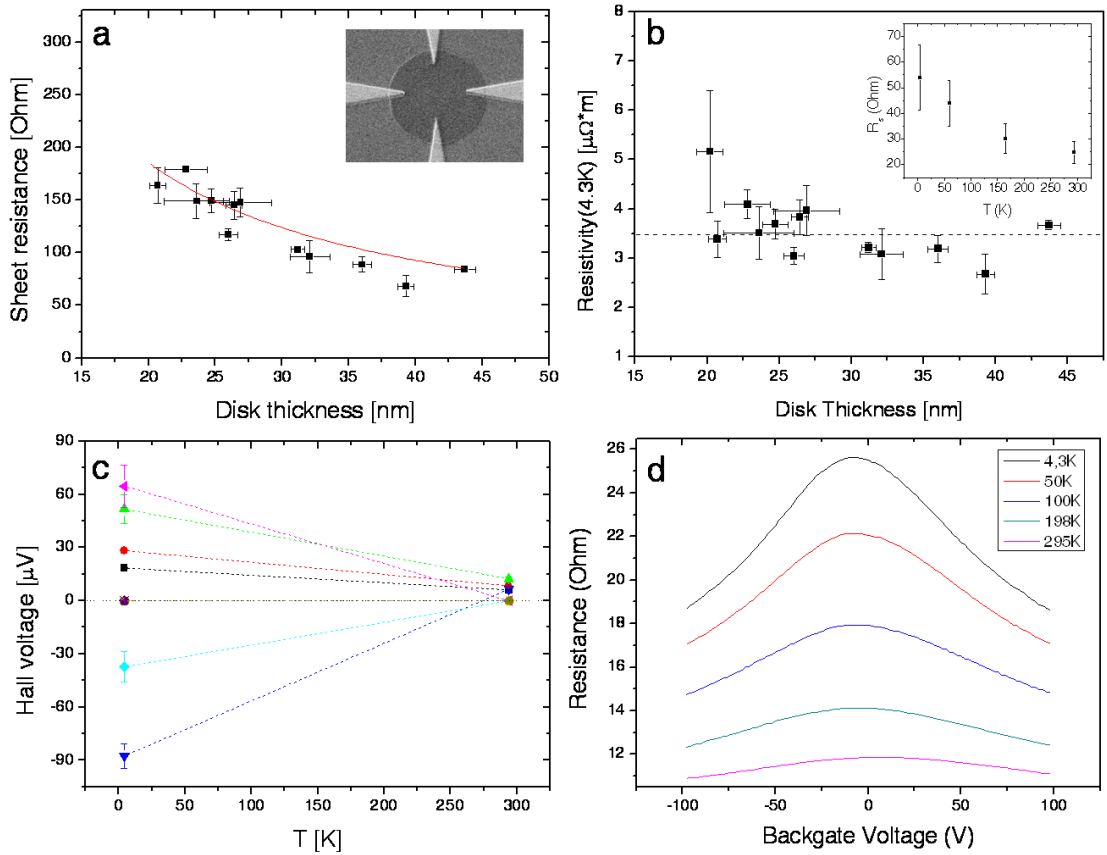

**Figure 3.** Transport measurements (a) Sheet resistance (Rs) vs. disk thickness at LT; the red curve is the reciprocal fitting of the experimental data: (inset) SEM image of a disk with contacts in a vdP configuration. (b) vdP corrected resistivity vs. disk thickness exhibiting an almost constant value; (inset) Temperature dependance of Rs for a disk. (c) vdP corrected Hall voltages as a function of temperature displaying a broad distribution of values at low temperature for different disks. (d) Back-gate dependence of the resistance for a single disk at different temperatures.

**Supporting Information**. Structural characterization: Atomic force microscopy, Transmission electron microscopy, Raman Spectroscopy, Kelvin Force Probe Microscopy. Transport properties: Calculation of the sheet resistance y temperature dependence of the resistance. This material is available free of charge via the Internet at http://pubs.acs.org.

**Corresponding Authors**

Correspondence should be addressed to M. K. Klaeui@uni-mainz.de and K. M. muellen@mpip-mainz.mpg.de

**ACKNOWLEDGMENTS**

This work was financially supported by the DFG (Priority Program Graphene SPP 1459, KL1811), the ERC ((ERC-2007-StG 208162) the European Science Foundation (ESF) under the EUROCORES Program EuroGRAPHENE (GOSPEL) and the EC Marie-Curie ITN-GENIUS (PITN-GA-2010-264694). Y.H. gratefully acknowledges funding by the Alexander von Humboldt Foundation.